\documentclass[aps,prd,reprint,superscriptaddress,nofootinbib]{revtex4-2}

\usepackage[utf8]{inputenc}
\usepackage[T1]{fontenc}
\usepackage[english]{babel}
\usepackage{amsmath}
\usepackage{amssymb}
\usepackage{booktabs}
\usepackage{graphicx}
\usepackage[caption=false]{subfig}
\usepackage{microtype}
\usepackage{xcolor}
\usepackage{underscore}
\DeclareUnicodeCharacter{2212}{-}

\usepackage{hyperref}
\hypersetup{
    colorlinks=true,
    linkcolor=blue,
    citecolor=blue,
    urlcolor=cyan,
    pdftitle={Chiral Couplings of a Light Lmu-Ltau-like Z': Reinterpretation of CMS Four-Muon Data and High-Luminosity LHC Projections},
    pdfauthor={Cesar Rendon}
}

\begin{document}

\title{Chiral Couplings of a Light {\boldmath$L_\mu - L_\tau$}-like {\boldmath$Z^\prime$}: Reinterpretation of CMS Four-Muon Data and High-Luminosity LHC Projections}

\author{C\'esar Rend\'on}
\email{cesar.rendon@ugent.be}
\affiliation{Department of Physics and Astronomy, Ghent University, Proeftuinstraat 86, B-9000 Ghent, Belgium}

\date{\today}

\begin{abstract}
We reinterpret the CMS search for a light $Z^\prime$ boson in $Z\to4\mu$ events with $77.3~\mathrm{fb}^{-1}$ at $\sqrt{s}=13$~TeV for a $Z^\prime$ with the flavor structure of $L_\mu-L_\tau$ and independent left- and right-handed couplings, parametrized by $\epsilon=g_R/g_L$. Signal events are simulated with \textsc{MadGraph5\_aMC@NLO}, \textsc{Pythia}~8 and \textsc{Delphes} and analyzed with the event selection of the CMS search, while the background is digitized from the same CMS publication. For the vector model ($\epsilon=1$) the expected limits agree with those of CMS within $9\%$ over $5$--$62$~GeV. For chiral couplings, $\epsilon=-1,3,7,\infty$, we separate the effects on the production cross section, the branching ratio and the acceptance. For the axial model ($\epsilon=-1$) the production cross section and the acceptance are the same as for the vector model. For $|g_L|\neq|g_R|$ the production cross section decreases by up to 15\% and the acceptance by up to 9\% at low masses, while the branching ratio increases. The expected limit on $\bar g=\sqrt{(g_L^2+g_R^2)/2}$ changes by at most $15\%$, and rescaling the vector limits with the ratio of branching ratios only overestimates the sensitivity by up to 12\%. For right-handed-dominated couplings, constraints from neutrino trident production, which require a coupling to neutrinos, no longer apply, while the four-muon channel remains sensitive through $g_R$. We project the sensitivity of the High-Luminosity LHC with $3000~\mathrm{fb}^{-1}$ at $\sqrt{s}=14$~TeV under several assumptions on the background systematic uncertainty and on pileup, reaching $\bar g\simeq1.5$--$2.1\times10^{-3}$ at $m_{Z^\prime}=5$~GeV.
\end{abstract}

\maketitle

\section{Introduction}
\label{sec:intro}

Gauge extensions of the Standard Model (SM) with an additional $U(1)$ symmetry predict a new neutral vector boson, $Z^\prime$, whose couplings to SM fermions depend on the charge assignments. A well-studied example is the gauged $U(1)_{L_\mu-L_\tau}$ symmetry~\cite{He:1990pn, He:1991qd}, which is anomaly free with the SM fermion content and in which the $Z^\prime$ couples only to the second- and third-generation leptons. Since the $Z^\prime$ does not couple to electrons or quarks at tree level, the main constraints on a light $Z^\prime$ of this type come from processes involving muons, taus and their neutrinos~\cite{Altmannshofer:2014pba, Bauer:2018onh}. The model was also proposed~\cite{Gninenko:2001hx} to explain the difference between the measured muon anomalous magnetic moment~\cite{Muong-2:2023xgm} and its SM prediction~\cite{Aoyama:2020ynm}. However, because the most recent SM prediction agrees with the measurement~\cite{Aliberti:2025wp}, we do not use $(g-2)_\mu$ to select a region of parameter space.

\subsection{Experimental landscape}
\label{sec:landscape}

At the LHC, the process $Z\to\mu^+\mu^-$ can produce a light $Z^\prime$ through radiation from one of the muons, $Z\to\mu^+\mu^-Z^\prime$, followed by $Z^\prime\to\mu^+\mu^-$. The resulting four-muon final state has a small SM background, and the signal appears as a narrow peak in the invariant mass of one of the dimuon pairs. The CMS Collaboration searched for this signature with $77.3~\mathrm{fb}^{-1}$ at $\sqrt{s}=13$~TeV~\cite{CMS:2018yxg}, and the ATLAS Collaboration with $139~\mathrm{fb}^{-1}$ at the same energy, using a parametrized neural network~\cite{ATLAS:2023fhg}.

Both searches interpret their results in terms of the purely vector current of the $L_\mu-L_\tau$ model, assuming left-right symmetric couplings (\(g_L = g_R\)). However, the $Z^\prime$ couplings to left- and right-handed leptons need not be equal. Chiral couplings appear in models with additional vector-like fermions~\cite{Ellis:2017nrp, King:2017anf}, and an axial version of $L_\mu-L_\tau$ with a renormalizable ultraviolet completion was recently proposed~\cite{Fang:2026axial}. In that work, the CMS four-muon limits were rescaled to the axial case using the ratio of branching ratios, based on the argument that, for $m_{Z^\prime}\gg m_\mu$, vector and axial radiation lead to nearly identical selection efficiencies. This argument does not apply to couplings with $|g_L|\neq|g_R|$, for which both the production cross section and the acceptance may depend on the chirality of the couplings.

Building on these searches, we reinterpret the CMS analysis~\cite{CMS:2018yxg} for a $Z^\prime$ with the flavor structure of $L_\mu-L_\tau$ and independent left- and right-handed couplings, parametrized by $\epsilon=g_R/g_L$. We use the CMS analysis because its selection is based on a mass window. While the ATLAS search sets stronger limits, the dependence of its sensitivity on $\epsilon$ requires a dedicated study. After validating the reinterpretation against the CMS expected limits for the vector case, we determine how the production cross section, the branching ratio and the acceptance depend on $\epsilon$ and derive the corresponding expected limits. We also present projections for the High-Luminosity LHC (HL-LHC).

The paper is organized as follows. Section~\ref{sec:model} describes the model. Sections~\ref{sec:recast} and~\ref{sec:selection} present the simulation, the event selection, the background and the statistical treatment, and the reinterpretation is validated in Sec.~\ref{sec:validation}. The results for chiral couplings are given in Sec.~\ref{sec:chiral}. The HL-LHC projections are presented in Sec.~\ref{sec:hllhc}, and the results are compared with other constraints in Sec.~\ref{sec:constraints}. Section~\ref{sec:conclusions} contains our conclusions, and the theoretical consistency of the model is discussed in Appendix~\ref{app:eft}.

\section{Model}
\label{sec:model}

\subsection{Chiral couplings with $L_\mu - L_\tau$ flavor structure}
\label{sec:lagrangian}

We consider a neutral vector boson $Z^\prime$ coupled to the second- and third-generation leptons with the flavor structure of $L_\mu-L_\tau$~\cite{He:1990pn, He:1991qd}, that is, with opposite charges for the muon and tau families and no tree-level couplings to electrons or quarks, but with independent couplings to left- and right-handed fields:
\begin{equation}
\begin{split}
\mathcal{L}_{Z^\prime} = Z^\prime_\alpha \Big[\,
  &\bar\mu\gamma^\alpha\big(g_L P_L + g_R P_R\big)\mu
  - \bar\tau\gamma^\alpha\big(g_L P_L + g_R P_R\big)\tau \\
  &+ g_L\,\bar\nu_\mu\gamma^\alpha P_L\nu_\mu
  - g_L\,\bar\nu_\tau\gamma^\alpha P_L\nu_\tau \Big],
\end{split}
\label{eq:lagrangian}
\end{equation}
where $P_{L,R}=(1\mp\gamma_5)/2$. The neutrinos couple with the same strength as the left-handed charged leptons, as required by $SU(2)_L$ invariance. We parametrize the chirality of the couplings by
\begin{equation}
\epsilon = \frac{g_R}{g_L}, \qquad \bar g = \sqrt{\frac{g_L^2+g_R^2}{2}} ,
\label{eq:epsilon}
\end{equation}
so that the purely vector $L_\mu-L_\tau$ model corresponds to $\epsilon=1$ with $\bar g=g$, the axial model of Ref.~\cite{Fang:2026axial} to $\epsilon=-1$, and $\epsilon\to\infty$ to a $Z^\prime$ coupled only to right-handed leptons. We study the benchmarks $\epsilon=-1,1,3,7,\infty$, listed in Table~\ref{tab:benchmarks}. For $\epsilon\neq\pm1$ the model is not anomaly free with the SM fermions alone; its theoretical consistency is discussed in Appendix~\ref{app:eft}.

\begin{table}[t]
\caption{Benchmark couplings used in the simulation. The couplings to taus have the opposite sign.}
\label{tab:benchmarks}
\begin{ruledtabular}
\begin{tabular}{lccc}
$\epsilon$ & $g_L$ & $g_R$ & $\bar g$ \\
\hline
$-1$      & 0.01000 & $-0.01000$ & 0.0100 \\
$1$       & 0.01000 & 0.01000    & 0.0100 \\
$3$       & 0.00447 & 0.01342    & 0.0100 \\
$7$       & 0.00200 & 0.01400    & 0.0100 \\
$\infty$  & 0       & 0.01414    & 0.0100 \\
\end{tabular}
\end{ruledtabular}
\end{table}

The model is implemented in \textsc{FeynRules}~\cite{ALLOUL20142250}, starting from the $W^\prime$ and $Z^\prime$ model of Refs.~\cite{Fuks:2017vtl, FeynRules:WZPrimeAtNLO}, which we modified to include independent left- and right-handed couplings to muons and taus, and exported in the UFO format~\cite{DEGRANDE20121201}.

\subsection{Decay widths and branching ratios}
\label{sec:widths}

For a charged lepton of mass $m_\ell$, with $x_\ell=m_\ell^2/m_{Z^\prime}^2$ and $\beta_\ell=\sqrt{1-4x_\ell}$, the partial widths are
\begin{align}
\Gamma(Z^\prime\to\ell^+\ell^-) &= \frac{m_{Z^\prime}}{24\pi}\,\beta_\ell\Big[(g_L^2+g_R^2)(1-x_\ell)+6\,g_Lg_R\,x_\ell\Big], \label{eq:gll}\\
\Gamma(Z^\prime\to\nu\bar\nu) &= \frac{g_L^2\,m_{Z^\prime}}{24\pi}\quad\text{per neutrino flavor}. \label{eq:gnn}
\end{align}

\begin{table*}[t]
\caption{Branching ratios of the $Z^\prime$ to $\mu^+\mu^-$ and $\tau^+\tau^-$ for all benchmarks, and partial width $\Gamma(Z^\prime\to\mu^+\mu^-)$, for $\bar g=0.01$, computed from Eqs.~\eqref{eq:gll}--\eqref{eq:gnn}. The branching ratio to neutrinos, summed over flavors, is $1-\mathcal B(\mu^+\mu^-)-\mathcal B(\tau^+\tau^-)$. For fixed $\bar g$, $\Gamma(Z^\prime\to\mu^+\mu^-)$ is independent of $\epsilon$ within 0.3\%, so that the dependence of $\mathcal B(Z^\prime\to\mu^+\mu^-)$ on $\epsilon$ arises from the total width. The values obtained with \textsc{MadGraph5\_aMC@NLO} agree within 0.03\%.}
\label{tab:br}
\begin{ruledtabular}
\begin{tabular}{rccccccccccc}
 & \multicolumn{5}{c}{$\mathcal B(Z^\prime\to\mu^+\mu^-)$} & \multicolumn{5}{c}{$\mathcal B(Z^\prime\to\tau^+\tau^-)$} & $\Gamma(Z^\prime\to\mu^+\mu^-)$ \\
$m_{Z^\prime}$ [GeV] & $\epsilon=-1$ & $1$ & $3$ & $7$ & $\infty$ & $\epsilon=-1$ & $1$ & $3$ & $7$ & $\infty$ & [$10^{-5}$ GeV] \\
\hline
5  & 0.425 & 0.347 & 0.506 & 0.578 & 0.619 & 0.148 & 0.306 & 0.392 & 0.399 & 0.381 & 1.33 \\
9  & 0.360 & 0.334 & 0.466 & 0.512 & 0.531 & 0.279 & 0.331 & 0.441 & 0.468 & 0.469 & 2.39 \\
13 & 0.346 & 0.334 & 0.459 & 0.500 & 0.514 & 0.308 & 0.333 & 0.449 & 0.480 & 0.486 & 3.45 \\
17 & 0.341 & 0.333 & 0.457 & 0.496 & 0.508 & 0.319 & 0.333 & 0.451 & 0.484 & 0.492 & 4.51 \\
22 & 0.338 & 0.333 & 0.456 & 0.494 & 0.505 & 0.325 & 0.333 & 0.453 & 0.487 & 0.495 & 5.84 \\
28 & 0.336 & 0.333 & 0.456 & 0.492 & 0.503 & 0.328 & 0.333 & 0.453 & 0.488 & 0.497 & 7.43 \\
34 & 0.335 & 0.333 & 0.455 & 0.492 & 0.502 & 0.330 & 0.333 & 0.454 & 0.489 & 0.498 & 9.02 \\
40 & 0.335 & 0.333 & 0.455 & 0.491 & 0.501 & 0.331 & 0.333 & 0.454 & 0.489 & 0.499 & 10.6 \\
50 & 0.334 & 0.333 & 0.455 & 0.491 & 0.501 & 0.332 & 0.333 & 0.454 & 0.490 & 0.499 & 13.3 \\
62 & 0.334 & 0.333 & 0.455 & 0.491 & 0.501 & 0.332 & 0.333 & 0.454 & 0.490 & 0.499 & 16.4 \\
\end{tabular}
\end{ruledtabular}
\end{table*}

Since the neutrinos couple only through $g_L$, $\mathcal{B}(Z^\prime\to\mu^+\mu^-)$ increases with $\epsilon$, from $1/3$ for $\epsilon=1$ to $1/2$ for $\epsilon\to\infty$ in the massless limit (Table~\ref{tab:br}). Near the $\tau^+\tau^-$ threshold, the $\tau^+\tau^-$ width is suppressed by the $\tau$ mass, which increases $\mathcal{B}(Z^\prime\to\mu^+\mu^-)$ at low masses for all benchmarks. The suppression is strongest for $\epsilon=-1$, where the term proportional to $g_Lg_R$ in Eq.~\eqref{eq:gll} is negative; at $m_{Z^\prime}=5$~GeV, $\mathcal{B}(Z^\prime\to\tau^+\tau^-)=0.148$, compared with 0.306 for $\epsilon=1$. As $m_{Z^\prime}$ increases, the effect of the $\tau$ mass becomes negligible, and the branching ratios approach their values in the massless limit. For all benchmarks $\Gamma_{Z^\prime}/m_{Z^\prime}<10^{-5}$, well below the detector resolution, so that the signal rate scales as $\bar g^2$ for fixed $\epsilon$.


\section{Simulation}
\label{sec:recast}

\subsection{Event generation}
\label{sec:generation}

Signal events for $pp\to\mu^+\mu^-\mu^+\mu^-$ are generated at leading order with \textsc{MadGraph5\_aMC@NLO}~v3.5.16~\cite{Alwall:2014hca}, using the model of Sec.~\ref{sec:model}. We include only the diagrams with $Z$-boson exchange and exactly one $Z^\prime$ propagator, in which the $Z^\prime$ is radiated from one of the muons and decays to a muon pair, as in the signal topology considered by CMS~\cite{CMS:2018yxg}. Diagrams with photon or Higgs-boson exchange, and the SM four-muon continuum, are not included in the process. The parton distribution functions are taken from the \texttt{NNPDF40\_lo\_as\_01180} set~\cite{NNPDF:2021njg} through \textsc{LHAPDF6}~\cite{Buckley:2014ana}, in the five-flavor scheme. Parton showering and hadronization are simulated with \textsc{Pythia}~8.317~\cite{Bierlich:2022rug}.

At generator level we require $p_T>4$~GeV for each muon and $\Delta R>0.001$ between muons, with no requirement on the dimuon invariant mass. These requirements are looser than the event selection of Sec.~\ref{sec:selection}; in particular, the default minimum $\Delta R$ would remove a large fraction of the collimated muon pairs from low-mass $Z^\prime$ decays. We generate $5\times10^4$ events at $\sqrt{s}=13$~TeV for each of the ten mass points $m_{Z^\prime}=5$, 9, 13, 17, 22, 28, 34, 40, 50 and 62~GeV and each benchmark of Table~\ref{tab:benchmarks}, with $\bar g=0.01$. The $Z^\prime$ width is computed for each point from the model parameters.

The expected limits on $\bar g$ are obtained by rescaling the samples generated with $\bar g=0.01$, because the signal rate scales as $\bar g^2$ for fixed $\epsilon$ (Sec.~\ref{sec:widths}). The signal cross sections are multiplied by the NNLO/LO $K$-factor of 1.29 used in the CMS analysis~\cite{CMS:2018yxg}.

\subsection{Detector simulation}
\label{sec:delphes}

The detector response is simulated with \textsc{Delphes}~3.5.1~\cite{deFavereau:2013fsa}, starting from its standard CMS configuration, with two modifications to the muon reconstruction to follow the CMS analysis~\cite{CMS:2018yxg}. First, the muon identification efficiency is set to 0.94 for $p_T>5$~GeV and $|\eta|<2.4$. Together with the tracking efficiency of the configuration (0.98--0.99), this gives a combined reconstruction and identification efficiency of about 92\% per muon, as reported by CMS for signal events~\cite{CMS:2018yxg}. The CMS search does not give the dependence of this efficiency on $p_T$ and $\eta$, and we assume it to be constant above 5~GeV. The muon identification efficiencies measured by CMS for $p_T>20$~GeV are nearly independent of $p_T$ and $\eta$~\cite{CMS:2018rim}; since they may be lower at low $p_T$, the effect of a lower efficiency for $5<p_T<10$~GeV is estimated in Sec.~\ref{sec:validation}. Second, the muon isolation is computed in a cone of $\Delta R=0.3$ from the particle-flow candidates, excluding electrons and muons, so that the other muons of the event do not enter the isolation sum. No isolation requirement is applied at this stage; the requirement of the CMS analysis is applied in the event selection. The momentum resolution of the standard configuration is kept; its adequacy is tested in Sec.~\ref{sec:validation}.

\subsection{Differences with respect to the CMS simulation}
\label{sec:differences}

Our simulation differs from that of CMS~\cite{CMS:2018yxg} in several aspects. The detector response is emulated with \textsc{Delphes} instead of a full simulation, and the muon efficiency is assumed to be constant above 5~GeV (Sec.~\ref{sec:delphes}). Pileup is not included; its effect is discussed in Sec.~\ref{sec:hl_pileup}. CMS assumes $\mathcal{B}(Z^\prime\to\mu^+\mu^-)=1/3$, while we use the branching ratio computed from Eqs.~\eqref{eq:gll}--\eqref{eq:gnn}. Because of the $\tau$ mass, for $\epsilon=1$ it is 4\% larger than $1/3$ at $m_{Z^\prime}=5$~GeV, and the difference decreases at higher masses. Given that the signal yield scales as $\bar g^{2}\,\mathcal{B}(Z^\prime\to\mu^+\mu^-)$, this lowers our expected limit on $\bar g$ by about 2\% at the lowest mass. We also use different generator versions and PDF set (\textsc{MadGraph5\_aMC@NLO} v3.5.16 with NNPDF4.0 and \textsc{Pythia}~8.317, instead of v2.4.2 with NNPDF3.0 and \textsc{Pythia}~8.212~\cite{CMS:2018yxg}). All these differences affect only the signal, because the background is taken from the CMS analysis (Sec.~\ref{sec:background}). Their combined effect is assessed in Sec.~\ref{sec:validation} by comparing our expected limits with those of CMS.

\section{Event selection and background}
\label{sec:selection}

\subsection{Event selection}
\label{sec:evsel}

The event selection follows the CMS analysis~\cite{CMS:2018yxg} and is applied to the reconstructed muons using the \textsc{Coffea}~\cite{Smith:2020Coffea} and \textsc{Awkward Array}~\cite{Pivarski:awkward} libraries.

Muons are required to have $p_T>5$~GeV, $|\eta|<2.4$ and a relative isolation $I^\mu<0.35$. The isolation $I^\mu$ is the scalar sum of the transverse momenta of the particle-flow candidates within $\Delta R<0.3$ of the muon, excluding other muons (Sec.~\ref{sec:delphes}), divided by the muon $p_T$. The same isolation requirement is used in the CMS $H\to ZZ\to4\ell$ analysis~\cite{CMS:2017jkd}. Events must contain at least four selected muons, at least two of them with $p_T>10$~GeV and at least one with $p_T>20$~GeV. These requirements imply the thresholds of the dimuon trigger used by CMS, $p_T>17$ and 8~GeV, and the trigger efficiency for events passing the full selection was measured by CMS to be above 99\%~\cite{CMS:2018yxg}. We therefore do not apply a trigger inefficiency.

Among the opposite-sign pairs of selected muons, the pair with the highest invariant mass is taken as the $Z_1$ candidate. The $Z_2$ candidate is formed from the remaining muons; if more than one opposite-sign pair is available, the pair with the highest scalar sum of $p_T$ is chosen. Events without two such pairs are rejected. The four muons of the candidate must then satisfy $\Delta R>0.02$ for all pairs, $m(Z_1)>12$~GeV, $m(\mu^+\mu^-)>4$~GeV for all opposite-sign pairs, and $80<m(4\mu)<100$~GeV. The requirement $m(\mu^+\mu^-)>4$~GeV rejects muons from low-mass resonances and from heavy-flavor hadron decays.

For each mass hypothesis, the signal region is defined as in the CMS analysis by a window of $\pm2\%$ around $m_{Z^\prime}$ in $m(Z_2)$ for $m_{Z^\prime}<42.65$~GeV, and in $m(Z_1)$ for higher masses. The fraction of selected signal events that falls within this window is compared with the CMS value in Sec.~\ref{sec:validation}.

\subsection{Background}
\label{sec:background}

The background is not simulated. We use the SM prediction of the CMS analysis~\cite{CMS:2018yxg}, which is dominated by resonant $Z$ production with $Z\to4\mu$, with smaller contributions from nonresonant $q\bar q\to4\mu$ and $gg\to4\mu$ production. The prediction and its systematic uncertainty are digitized with \textsc{WebPlotDigitizer}~\cite{WebPlotDigitizer} from the $m(Z_2)$ distribution for $4<m<42$~GeV and the $m(Z_1)$ distribution for $42<m<72$~GeV shown in Fig.~4 of Ref.~\cite{CMS:2018yxg}. In these distributions the bin width follows the expected mass resolution, $w=0.02\,m$ below 42~GeV and $w=0.01\,m$ above. The digitized points are interpolated linearly, and the expected number of background events in the signal window of width $0.04\,m_{Z^\prime}$ is
\begin{equation}
N_b(m_{Z^\prime}) = \bar h \, \frac{0.04\,m_{Z^\prime}}{w(m_{Z^\prime})},
\label{eq:nbkg}
\end{equation}
where $\bar h$ is the average height of the distribution within the window. The systematic uncertainty is obtained in the same way from the digitized uncertainty band.

To validate this procedure, we compare in Table~\ref{tab:bkg} our background estimate with the values given by CMS for three signal windows. The agreement is within 2\%, well below the uncertainties of the CMS prediction.

\begin{table}[t]
\caption{Expected background in three signal windows, from the digitized CMS distributions [Eq.~\eqref{eq:nbkg}] and as given by CMS~\cite{CMS:2018yxg}, with statistical and systematic uncertainties.}
\label{tab:bkg}
\begin{ruledtabular}
\begin{tabular}{lcc}
Window & This work & CMS \\
\hline
$4.9<m(Z_2)<5.1$~GeV   & 9.1  & $9.2\pm3.0\pm0.7$ \\
$14.7<m(Z_2)<15.3$~GeV & 7.6  & $7.7\pm2.8\pm0.6$ \\
$68.6<m(Z_1)<71.4$~GeV & 35.2 & $34.9\pm5.9\pm2.8$ \\
\end{tabular}
\end{ruledtabular}
\end{table}

\subsection{Statistical analysis}
\label{sec:stats}

For each mass hypothesis, the signal window of Sec.~\ref{sec:evsel} is treated as a single-bin counting experiment with an integrated luminosity of $77.3~\mathrm{fb}^{-1}$. The likelihood is built with \textsc{pyhf}~\cite{Heinrich:2021pyhf, Heinrich:2022chep} from the expected signal yield $\mu N_s$, where $N_s$ is the yield for $\bar g=0.01$ and $\mu$ is the signal strength, and the expected background $N_b$ of Eq.~\eqref{eq:nbkg}. The systematic uncertainty of the background is included as a nuisance parameter with a Poisson constraint. The likelihood is minimized with \textsc{iminuit}~\cite{Dembinski:2020iminuit}.

Upper limits on $\mu$ at 95\% confidence level are obtained with the CL$_s$ criterion, using the profile-likelihood test statistic in the asymptotic approximation~\cite{Cowan:2010js}. We compute expected limits from the background-only Asimov dataset, together with their $\pm1\sigma$ and $\pm2\sigma$ bands. Since the signal yield scales as $\bar g^2$ for fixed $\epsilon$, the limit on $\mu$ is converted into a limit on the coupling as $\bar g_{\mathrm{lim}}=0.01\sqrt{\mu_{\mathrm{lim}}}$.

\section{Validation against CMS}
\label{sec:validation}

\begin{figure}[t]
\centering
\includegraphics[width=\columnwidth]{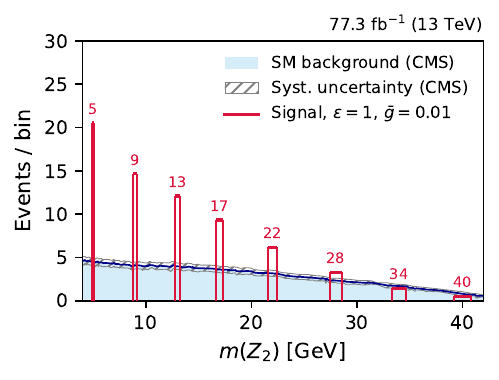}
\caption{Expected signal for $\epsilon=1$ and $\bar g=0.01$ in $m(Z_2)$ for $m_{Z^\prime}<42.65$~GeV, compared with the SM background predicted by CMS~\cite{CMS:2018yxg}. The numbers indicate $m_{Z^\prime}$ in GeV. The signal is shown as the average number of events per bin within the $\pm2\%$ signal window, using the variable bin width of the CMS distribution.}
\label{fig:mass_low}
\end{figure}

\begin{figure}[t]
\centering
\includegraphics[width=\columnwidth]{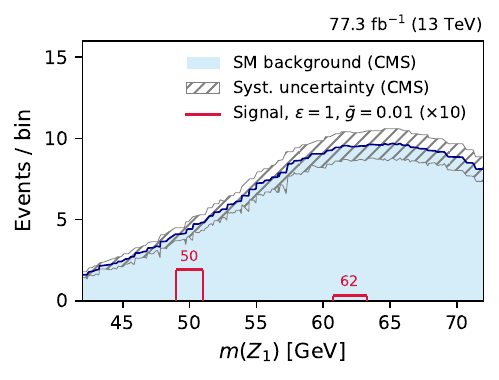}
\caption{Same as Fig.~\ref{fig:mass_low}, in $m(Z_1)$ for $m_{Z^\prime}>42.65$~GeV. The signal is multiplied by 10.}
\label{fig:mass_high}
\end{figure}

We validate the reinterpretation for the vector model, $\epsilon=1$, for which the CMS results are available~\cite{CMS:2018yxg}. Figures~\ref{fig:mass_low} and~\ref{fig:mass_high} show the expected signal for $\bar g=0.01$ in the $m(Z_2)$ and $m(Z_1)$ distributions, respectively, together with the SM background predicted by CMS. For $m_{Z^\prime}=5$~GeV, 41 signal events are expected in the signal window, compared with a background of 9.1 events. The signal yield decreases with $m_{Z^\prime}$, following the decrease of the production cross section and the mass dependence of the selection efficiency discussed below. Above 42.65~GeV, where the $Z^\prime$ is reconstructed as $Z_1$, the signal is much smaller than the background, which reaches its maximum near 65~GeV; the corresponding limits are therefore considerably weaker.

\begin{table}[t]
\caption{Fraction of selected signal events within the $\pm2\%$ signal window for $\epsilon=1$, compared with the values reported by CMS~\cite{CMS:2018yxg}.}
\label{tab:fwin}
\begin{ruledtabular}
\begin{tabular}{lccc}
$m_{Z^\prime}$ [GeV] & Observable & This work & CMS \\
\hline
5  & $m(Z_2)$ & 0.63 & 0.63 \\
40 & $m(Z_2)$ & 0.25 & 0.25 \\
\end{tabular}
\end{ruledtabular}
\end{table}

\begin{table*}[t]
\caption{Results for $\epsilon=1$. For each $m_{Z^\prime}$ we give the LO cross section for $\bar g=0.01$, the efficiency of the full selection $\varepsilon_{\mathrm{sel}}$, the fraction of selected events within the $\pm2\%$ signal window $f_{\mathrm{win}}$, the expected numbers of signal ($N_s$, for $\bar g=0.01$ and including the $K$-factor) and background ($N_b$) events in the window for $77.3~\mathrm{fb}^{-1}$, the systematic uncertainty of the background $\delta N_b$, and the expected 95\% CL upper limit on $\bar g$ with its $\pm1\sigma$ and $\pm2\sigma$ ranges.}
\label{tab:eps1}
\begin{ruledtabular}
\begin{tabular}{rcccrrrccc}
$m_{Z^\prime}$ [GeV] & $\sigma_{\mathrm{LO}}$ [fb] & $\varepsilon_{\mathrm{sel}}$ & $f_{\mathrm{win}}$ & $N_s$ & $N_b$ & $\delta N_b$ & $\bar g_{\mathrm{lim}}$ ($10^{-3}$) & $\pm1\sigma$ ($10^{-3}$) & $\pm2\sigma$ ($10^{-3}$) \\
\hline
5 & 1.98 & 0.332 & 0.63 & 41.1 & 9.1 & 0.83 & 4.25 & [3.51, 5.19] & [2.98, 6.26] \\
9 & 1.40 & 0.346 & 0.61 & 29.3 & 8.2 & 0.79 & 4.92 & [4.06, 6.02] & [3.44, 7.27] \\
13 & 1.09 & 0.362 & 0.62 & 24.2 & 8.0 & 0.77 & 5.38 & [4.45, 6.59] & [3.77, 7.97] \\
17 & 0.812 & 0.384 & 0.60 & 18.7 & 7.2 & 0.75 & 6.02 & [4.96, 7.38] & [4.20, 8.93] \\
22 & 0.540 & 0.410 & 0.56 & 12.3 & 6.4 & 0.68 & 7.21 & [5.93, 8.86] & [5.02, 10.8] \\
28 & 0.311 & 0.435 & 0.49 & 6.57 & 4.7 & 0.52 & 9.30 & [7.62, 11.5] & [6.42, 14.1] \\
34 & 0.169 & 0.448 & 0.39 & 2.91 & 3.4 & 0.40 & 13.2 & [10.7, 16.4] & [9.01, 20.2] \\
40 & 0.0884 & 0.454 & 0.25 & 1.01 & 1.6 & 0.47 & 19.8 & [15.9, 25.1] & [13.3, 31.3] \\
50 & 0.0271 & 0.422 & 0.67 & 0.763 & 17.7 & 1.94 & 36.6 & [30.5, 44.1] & [26.1, 52.5] \\
62 & 0.00513 & 0.311 & 0.81 & 0.129 & 38.0 & 3.45 & 107 & [90.1, 128] & [77.3, 151] \\
\end{tabular}
\end{ruledtabular}
\end{table*}

Table~\ref{tab:eps1} summarizes the results for $\epsilon=1$. The efficiency of the full selection, before the requirement on the signal window, increases from 33\% at $m_{Z^\prime}=5$~GeV to 45\% at 40~GeV, and decreases to 31\% at 62~GeV. The fraction of selected events within the $\pm2\%$ signal window is compared with the CMS values in Table~\ref{tab:fwin}. It is 63\% at 5~GeV and 25\% at 40~GeV, in agreement with CMS. The lower value at 40~GeV is due to the ambiguity in the assignment of the muons to the $Z_1$ and $Z_2$ candidates near $m_Z/2$~\cite{CMS:2018yxg}. Since this fraction depends on the momentum resolution and on the pairing of the muons, the agreement indicates that both are well reproduced by the simulation.

\begin{figure}[t]
\centering
\includegraphics[width=\columnwidth]{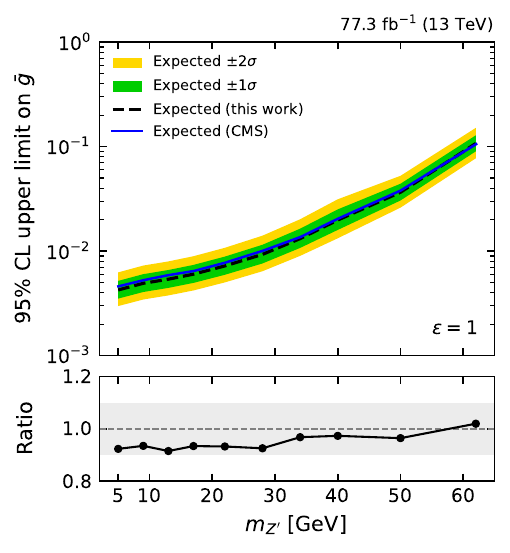}
\caption{Expected 95\% CL upper limits on $\bar g$ for $\epsilon=1$ as a function of $m_{Z^\prime}$, with the $\pm1\sigma$ and $\pm2\sigma$ bands, compared with the expected limits of CMS~\cite{CMS:2018yxg}. The lower panel shows the ratio of our expected limits to those of CMS; the gray band indicates $\pm10\%$.}
\label{fig:limits_cms}
\end{figure}

Figure~\ref{fig:limits_cms} compares our expected limits for $\epsilon=1$ with those of CMS~\cite{CMS:2018yxg}. The two agree within 9\% over the full mass range, well within the $\pm1\sigma$ band of the expected limit. For $m_{Z^\prime}\le28$~GeV our limits are 6--9\% stronger than those of CMS, while for higher masses the difference is below 4\%. Part of the difference at low masses comes from the branching ratio: CMS assumes $\mathcal{B}(Z^\prime\to\mu^+\mu^-)=1/3$, while our value is 4\% larger at 5~GeV (Sec.~\ref{sec:differences}), which lowers our limit by about 2\%. The remaining difference is consistent with the simplifications of our simulation that affect mainly low-$p_T$ muons, which are more frequent at low $m_{Z^\prime}$: the constant muon efficiency above 5~GeV and the absence of pileup, which both tend to overestimate the signal efficiency. Reducing the muon efficiency to 0.85 for $5<p_T<10$~GeV lowers the signal yield by 10\% and weakens our expected limit by 6\% at both $m_{Z^\prime}=5$ and 9~GeV, which reduces the difference with CMS from 8\% to 3\% and from 7\% to 1\%, respectively. Uncertainties in the signal yield, which are included by CMS, are not included in our statistical model; including a 10\% uncertainty weakens the expected limits by less than 1\%. We conclude that the reinterpretation reproduces the CMS sensitivity to better than 10\%, which is sufficient for the study of the chiral couplings in Sec.~\ref{sec:chiral}, where the results are given relative to $\epsilon=1$.


\section{Chiral couplings}
\label{sec:chiral}

We now apply the analysis of Secs.~\ref{sec:recast} and~\ref{sec:selection} without modification to the benchmarks $\epsilon=-1$, 3, 7 and $\infty$ of Table~\ref{tab:benchmarks}. All results are given for $\bar g=0.01$ at generation, and are compared with those for $\epsilon=1$.

\subsection{Production, branching ratio and acceptance}
\label{sec:decomposition}

For a fixed mass and fixed $\bar g$, the expected number of signal events relative to the vector model can be written as
\begin{equation}
\frac{N_s(\epsilon)}{N_s(1)}=P(\epsilon)\,\frac{\mathcal B(\epsilon)}{\mathcal B(1)}\,\frac{A(\epsilon)}{A(1)},
\label{eq:decomp}
\end{equation}
where $P(\epsilon)$ is the ratio of the production cross sections of $pp\to\mu^+\mu^-Z^\prime$, $\mathcal B(\epsilon)=\mathcal B(Z^\prime\to\mu^+\mu^-)$, and $A(\epsilon)$ is the fraction of generated events that pass the selection and fall within the signal window. The factorization of the production and the decay holds because the $Z^\prime$ width is much smaller than its mass: $\Gamma_{Z^\prime}/m_{Z^\prime}<10^{-5}$ for $\bar g=0.01$ (Sec.~\ref{sec:widths}), and below $10^{-3}$ at the excluded couplings.

The dependence on $\epsilon$ follows from the conservation of chirality along a massless muon line: the $Z$ and $Z^\prime$ vertices do not mix left- and right-handed fields, and chirality-flipping effects are suppressed by $m_\mu^2/m_{Z^\prime}^2<5\times10^{-4}$. The muon line produced by the $Z$ and the line from the $Z^\prime$ decay therefore have definite chiralities $a$ and $b$, and contributions with different $(a,b)$ do not interfere. Denoting the $Z$ couplings to left- and right-handed muons by $c_L=-1/2+s_W^2$ and $c_R=s_W^2$, the squared matrix element is a sum of four terms with weights $c_a^2g_a^2g_b^2$ and kinematic distributions that do not depend on $\epsilon$. For identical muons, the interference between exchanged diagrams contributes only for $a=b$, with the same weights.

Summing over the decay chirality, and using that the integrated rates for left- and right-handed muon lines are equal for the forward-backward symmetric $pp$ initial state and selection (Sec.~\ref{sec:selection}), the production ratio is
\begin{equation}
P(\epsilon)=\frac{c_L^2g_L^2+c_R^2g_R^2}{(c_L^2+c_R^2)\,\bar g^2}
=\frac{(1+\mathcal{A}_\ell)+\epsilon^2(1-\mathcal{A}_\ell)}{1+\epsilon^2},
\label{eq:P}
\end{equation}
where $\mathcal{A}_\ell=(c_L^2-c_R^2)/(c_L^2+c_R^2)\simeq0.15$ is the lepton asymmetry parameter of the $Z$~\cite{ParticleDataGroup:2024cfk}. The $Z$ produces more left- than right-handed muons, and as a result a $Z^\prime$ coupled mainly to right-handed muons is radiated less often. Equation~\eqref{eq:P} gives $P=1$ for $\epsilon=\pm1$, and $P=0.879$, 0.855 and 0.849 for $\epsilon=3$, 7 and $\infty$. The values obtained from the simulated cross sections and the branching ratios of Sec.~\ref{sec:widths} are 0.998--1.004, 0.879--0.884, 0.855--0.859 and 0.850--0.855 over the ten mass points, in agreement with Eq.~\eqref{eq:P} and independent of the mass. The branching ratio increases with $\epsilon$ (Table~\ref{tab:br}), and for $\epsilon>1$ this increase is larger than the relative decrease of the production cross section, $P(\epsilon)$, so that the number of signal events increases by 20\% for $\epsilon=3$ and by about 27\% for $\epsilon\ge7$ at high masses, where the acceptance does not depend on $\epsilon$.

\begin{figure}[t]
\centering
\includegraphics[width=\columnwidth]{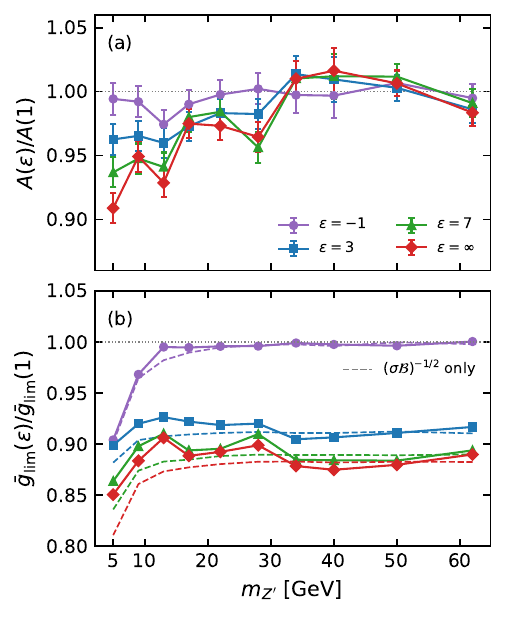}
\caption{(a) Signal acceptance for $\epsilon=-1$, 3, 7 and $\infty$ relative to $\epsilon=1$, with the statistical uncertainties of the simulated samples. (b) Expected limits on $\bar g$ relative to $\epsilon=1$ (solid lines), compared with the ratio obtained from the production cross section and the branching ratio only, $[\sigma(\epsilon)\mathcal B(\epsilon)/\sigma(1)\mathcal B(1)]^{-1/2}$ (dashed lines).}
\label{fig:chiral_ratios}
\end{figure}

Figure~\ref{fig:chiral_ratios}(a) shows the acceptance relative to $\epsilon=1$. For $\epsilon=-1$ the weights $c_a^2g_a^2g_b^2$ are the same as for $\epsilon=1$, since they depend only on $g_L^2$ and $g_R^2$, and the acceptance is equal to that of the vector model within the statistical uncertainty of the simulated samples, which is 1--2\% per mass point. This confirms, for the $Z\to4\mu$ channel and $m_{Z^\prime}\geq5$~GeV, the assumption of Ref.~\cite{Fang:2026axial}, where the limits for the axial model were obtained by rescaling the vector limits with the ratio of branching ratios. 

For $\epsilon=3$, 7 and $\infty$, the acceptance decreases at low masses, by 4\%, 6\% and 9\% at $m_{Z^\prime}=5$~GeV, and is compatible with that of the vector model above about 30~GeV. For these benchmarks, the fraction of events in which the $Z^\prime$ is emitted and decays through the same chirality,
\begin{equation}
f_{\mathrm{same}}=\frac{c_L^2g_L^4+c_R^2g_R^4}{\left(c_L^2g_L^2+c_R^2g_R^2\right)\left(g_L^2+g_R^2\right)},
\label{eq:fsame}
\end{equation}
increases from 0.5 for $\epsilon=\pm1$ to 0.80, 0.95 and 1. We denote by $\Delta(\epsilon)=1-A(\epsilon)/A(1)$ the relative decrease of the acceptance with respect to the vector model. If the acceptance depends only on whether the emission and the decay proceed through the same chirality, $A(\epsilon)=f_{\mathrm{same}}A_{\mathrm{same}}+(1-f_{\mathrm{same}})A_{\mathrm{opp}}$, and $\Delta(\epsilon)$ is proportional to $f_{\mathrm{same}}-1/2$. The ratios $\Delta(3)/\Delta(\infty)$ and $\Delta(7)/\Delta(\infty)$ are then expected to be 0.59 and 0.91. Averaging $\Delta(\epsilon)$ over 5--13~GeV, with the acceptances of Table~\ref{tab:cutflow}, the measured ratios are $0.53\pm0.08$ and $0.82\pm0.09$, consistent with this expectation.

The decrease of the acceptance comes from two steps of the selection: the rejection of events with an opposite-sign pair of mass below 4~GeV, whose efficiency at 5~GeV decreases from 0.890 for $\epsilon=1$ to 0.857 for $\epsilon=\infty$, and the fraction of selected events within the signal window, which decreases from 0.626 to 0.585. Both differences correspond to more than seven standard deviations of the simulated samples, and together they account for the observed change of the acceptance. The efficiencies of all other selection steps, including the trigger and the requirements on the muon transverse momenta, are the same for all values of $\epsilon$ (Appendix~\ref{app:cutflow}). We interpret this effect as follows. Both requirements involve opposite-sign pairs formed by a muon from the $Z^\prime$ decay and a muon from the $Z$ decay. At low $m_{Z^\prime}$, the $Z^\prime$ is emitted nearly collinear with the radiating muon, and whether the nearby $Z^\prime$ daughter has the opposite or the same charge as that muon depends on whether the emission and the decay proceed through the same chirality. A nearby pair of opposite charge can have a mass below 4~GeV, or modify the assignment of the muons to the $Z_1$ and $Z_2$ candidates. At higher masses these pairs have larger masses and the effect is reduced. At $m_{Z^\prime}=40$~GeV the efficiency of the veto decreases by only 1\% from $\epsilon=1$ to $\epsilon=\infty$, and the total acceptance is the same for all benchmarks within the statistical uncertainty (Table~\ref{tab:cutflow}).

\subsection{Expected limits as a function of $\epsilon$}
\label{sec:eps_limits}

The expected limits for all benchmarks are given in Table~\ref{tab:eps_limits}. As the signal region is a single-bin counting experiment with the same background for all $\epsilon$, the number of excluded signal events does not depend on $\epsilon$, and the ratio of the limits is given exactly by
\begin{equation}
\frac{\bar g_{\mathrm{lim}}(\epsilon)}{\bar g_{\mathrm{lim}}(1)}=\left[\frac{N_s(\epsilon)}{N_s(1)}\right]^{-1/2}.
\label{eq:glim_ratio}
\end{equation}
This ratio is shown in Fig.~\ref{fig:chiral_ratios}(b). For $\epsilon=3$, 7 and $\infty$ the limits are 7--15\% stronger than for the vector model over the full mass range. For $\epsilon=-1$ the limit is 10\% stronger at 5~GeV and within 2\% of the vector limit above 13~GeV: production and acceptance are the same as for $\epsilon=1$, and the difference comes only from the branching ratio, through the term proportional to $g_Lg_R$ in the $\tau^+\tau^-$ width near threshold [Eq.~\eqref{eq:gll}]. More generally, $P(\epsilon)$ and $A(\epsilon)$ depend only on $\epsilon^2$, because the weights $c_a^2g_a^2g_b^2$ contain only squared couplings, so that models with $\epsilon$ and $-\epsilon$ differ only through the branching ratio, up to corrections of order $m_\mu^2/m_{Z^\prime}^2$. Given that the efficiencies of the muon requirements do not depend on $\epsilon$, the effect of the low-$p_T$ muon efficiency discussed in Sec.~\ref{sec:validation} cancels in these ratios.

\begin{table}[t]
\caption{Expected 95\% CL upper limits on $\bar g$, in units of $10^{-3}$, for the five benchmarks.}
\label{tab:eps_limits}
\begin{ruledtabular}
\begin{tabular}{rccccc}
$m_{Z^\prime}$ [GeV] & $\epsilon=-1$ & $\epsilon=1$ & $\epsilon=3$ & $\epsilon=7$ & $\epsilon=\infty$ \\
\hline
5 & 3.84 & 4.25 & 3.82 & 3.67 & 3.61 \\
9 & 4.76 & 4.92 & 4.52 & 4.41 & 4.35 \\
13 & 5.36 & 5.38 & 4.99 & 4.90 & 4.88 \\
17 & 5.98 & 6.02 & 5.55 & 5.38 & 5.35 \\
22 & 7.18 & 7.21 & 6.62 & 6.45 & 6.43 \\
28 & 9.26 & 9.30 & 8.55 & 8.46 & 8.36 \\
34 & 13.2 & 13.2 & 11.9 & 11.6 & 11.6 \\
40 & 19.8 & 19.8 & 18.0 & 17.5 & 17.3 \\
50 & 36.4 & 36.6 & 33.3 & 32.3 & 32.2 \\
62 & 107 & 107 & 98.3 & 95.8 & 95.4 \\
\end{tabular}
\end{ruledtabular}
\end{table}

\subsection{Comparison with rate rescaling}
\label{sec:rescaling}

Limits for models not considered by an experimental analysis are often obtained by rescaling the published limits with the ratio of signal rates. Table~\ref{tab:decomp} compares our results with two such rescalings: with the ratio of branching ratios only, which is exact for the axial couplings considered in Ref.~\cite{Fang:2026axial}, where $P=1$, and with the ratio of production cross sections times branching ratios, which corresponds to the dashed lines of Fig.~\ref{fig:chiral_ratios}(b).

For $\epsilon=-1$ both rescalings reproduce our results within 1.5\%, since $P=1$ and the acceptance is unchanged. For couplings with $|g_L|\neq|g_R|$, extending the branching-ratio rescaling beyond the axial case gives limits that are too strong by 5--10\% above 15~GeV and by up to 12\% at 5~GeV, as it neglects the decrease of the production cross section, $P(\epsilon)<1$. Including $P(\epsilon)$, the rescaled limits agree with ours within about 2\% above 15~GeV, but are still too strong by 2--5\% between 5 and 13~GeV, where the acceptance decreases (Sec.~\ref{sec:decomposition}).

The factor $P(\epsilon)$ depends only on the $Z$ couplings to muons and applies to any search for $Z\to\mu^+\mu^-Z^\prime$ at the $Z$ pole. Limits from the ATLAS search~\cite{ATLAS:2023fhg} can therefore be estimated for chiral couplings by rescaling with $P(\epsilon)\,\mathcal B(\epsilon)/\mathcal B(1)$, with an additional acceptance correction of up to 9\% at low masses that depends on the event selection. In searches for $e^+e^-\to\mu^+\mu^-Z^\prime$ at $\sqrt s\approx10$~GeV, the muons are produced by a virtual photon with equal couplings to both chiralities, so that $P(\epsilon)=1$ up to corrections of order $s/m_Z^2$. The fraction of events with the same chirality in the emission and the decay of the $Z^\prime$ is, however, still larger than 1/2 for $|g_L|\neq|g_R|$, and spin-correlation effects on the acceptance analogous to those of Fig.~\ref{fig:chiral_ratios}(a) may remain.

\begin{table}[t]
\caption{Decomposition of the signal yield relative to $\epsilon=1$ [Eq.~\eqref{eq:decomp}] and ratio of the expected limits, $R_g=\bar g_{\mathrm{lim}}(\epsilon)/\bar g_{\mathrm{lim}}(1)$, compared with the ratios obtained by rescaling with $P\,\mathcal B$ and with $\mathcal B$ only. The statistical uncertainty of the acceptance ratios is 1--2\%.}
\label{tab:decomp}
\begin{ruledtabular}
\begin{tabular}{lcccccc}
$\epsilon$ & $P$ & $\mathcal B/\mathcal B_1$ & $A(\epsilon)/A_1$ & $R_g$ & $(P\mathcal B/\mathcal B_1)^{-1/2}$ & $(\mathcal B/\mathcal B_1)^{-1/2}$ \\
\hline
\multicolumn{7}{c}{$m_{Z^\prime}=5$~GeV} \\
$-1$     & 1.004 & 1.225 & 0.994 & 0.904 & 0.902 & 0.903 \\
$3$      & 0.881 & 1.459 & 0.963 & 0.899 & 0.882 & 0.828 \\
$7$      & 0.859 & 1.665 & 0.937 & 0.864 & 0.836 & 0.775 \\
$\infty$ & 0.853 & 1.783 & 0.909 & 0.851 & 0.811 & 0.749 \\
\hline
\multicolumn{7}{c}{$m_{Z^\prime}=22$~GeV} \\
$-1$     & 0.998 & 1.013 & 0.998 & 0.996 & 0.995 & 0.994 \\
$3$      & 0.881 & 1.368 & 0.983 & 0.919 & 0.911 & 0.855 \\
$7$      & 0.856 & 1.481 & 0.984 & 0.895 & 0.888 & 0.822 \\
$\infty$ & 0.852 & 1.515 & 0.973 & 0.892 & 0.880 & 0.813 \\
\hline
\multicolumn{7}{c}{$m_{Z^\prime}=62$~GeV} \\
$-1$     & 1.003 & 1.002 & 0.995 & 1.000 & 0.998 & 0.999 \\
$3$      & 0.884 & 1.364 & 0.986 & 0.917 & 0.911 & 0.856 \\
$7$      & 0.858 & 1.472 & 0.991 & 0.894 & 0.890 & 0.824 \\
$\infty$ & 0.855 & 1.502 & 0.984 & 0.890 & 0.883 & 0.816 \\
\end{tabular}
\end{ruledtabular}
\end{table}

\section{High-Luminosity LHC projections}
\label{sec:hllhc}

\subsection{Scaling to $\sqrt{s}=14$~TeV and $3000~\mathrm{fb}^{-1}$}
\label{sec:hl_scaling}

We project the sensitivity of the analysis to the High-Luminosity LHC, with an integrated luminosity of $3000~\mathrm{fb}^{-1}$ at $\sqrt s=14$~TeV. The event selection and the signal efficiencies are assumed to be the same as in the Run~2 analysis, and the assumptions on the background uncertainty are discussed in Sec.~\ref{sec:hl_syst}. As in Ref.~\cite{Cepeda:2019vtg}, we assume that the upgrades of the detectors compensate for the effects of the higher instantaneous luminosity and pileup on the object reconstruction. The signal and background yields are scaled with the luminosity and with the ratio of the cross sections at 14 and 13~TeV, computed with \textsc{MadGraph5\_aMC@NLO} with the same generator-level requirements. For the signal this ratio is 1.068--1.078 for $m_{Z^\prime}=5$--62~GeV, and we use the average value of 1.073. For the background, the SM $q\bar q\to4\mu$ cross section with requirements close to the event selection ($p_T>4$~GeV, $|\eta|<2.4$, $m(\mu^+\mu^-)>4$~GeV, $80<m(4\mu)<100$~GeV) increases by a factor of 1.072. Given that both the signal and the dominant background are produced through the $Z$ resonance in $q\bar q$ annihilation, the signal-to-background ratio is nearly unchanged, and the improvement comes from the increase of the luminosity. With $3000~\mathrm{fb}^{-1}$, the expected background in the signal window ranges from about 70 to 1600 events, so that the systematic uncertainty of the background becomes important.

\subsection{Systematic uncertainty scenarios}
\label{sec:hl_syst}

The background uncertainty at the HL-LHC is not known. In the HL-LHC projections of Ref.~\cite{Cepeda:2019vtg}, a scenario with the systematic uncertainties of the Run~2 analyses (S1) is compared with a scenario in which the experimental uncertainties are reduced according to the expected improvements and the theoretical uncertainties are halved (S2). Following this approach, we consider the relative background uncertainty of the Run~2 analysis multiplied by 1.0, 0.75 and 0.5, and a scenario with statistical uncertainties only. We take the reduction by a factor of 0.5 as the nominal scenario. The expected limits for $\epsilon=1$ are shown in Fig.~\ref{fig:hllhc}(a). In the nominal scenario, $\bar g$ values down to $1.7\times10^{-3}$ at $m_{Z^\prime}=5$~GeV and $5.4\times10^{-2}$ at 62~GeV can be probed, which improves the Run~2 limits by a factor of 2.5--2.7 below 40~GeV and of 2.0--2.1 at 50 and 62~GeV. The improvement ranges from 1.5--2.3 with the Run~2 uncertainty to 2.8--3.1 with statistical uncertainties only. The dependence on the systematic uncertainty is strongest at high masses, where the background is largest. Since the background is the same for all values of $\epsilon$, the ratios of the limits given in Sec.~\ref{sec:eps_limits} also apply to the HL-LHC projections.

\subsection{Impact of pileup}
\label{sec:hl_pileup}

\begin{figure}[t]
\centering
\includegraphics[width=\columnwidth]{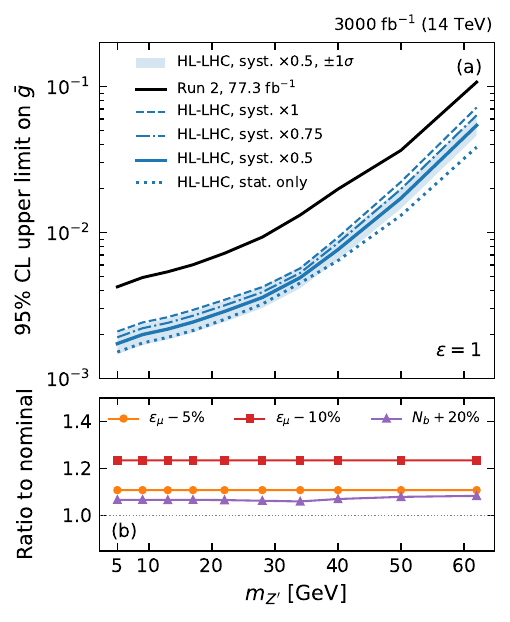}
\caption{(a) Expected 95\% CL upper limits on $\bar g$ for $\epsilon=1$ at the HL-LHC, $3000~\mathrm{fb}^{-1}$ at $\sqrt s=14$~TeV, for different assumptions on the background systematic uncertainty, relative to that of the Run~2 analysis, compared with the Run~2 expected limits. The band shows the $\pm1\sigma$ range of the nominal scenario, in which the uncertainty is reduced by a factor of 0.5. (b) Limits for variations of the muon efficiency and of the background that represent the effect of pileup, relative to the nominal scenario.}
\label{fig:hllhc}
\end{figure}

At the HL-LHC an average of up to 200 pp interactions per bunch crossing is expected, compared with about 30 in Run~2. Pileup is not included in our simulation (Sec.~\ref{sec:differences}). It mainly affects the muon isolation, and therefore the muon efficiency, and may increase the background from misidentified or nonprompt muons. We estimate its impact with three variations of the nominal scenario: a reduction of the efficiency per muon by 5\% and by 10\%, which reduces the signal yield by 19\% and 34\%, and an increase of the background by 20\%. Figure~\ref{fig:hllhc}(b) shows the resulting limits relative to the nominal scenario. The efficiency reductions weaken the limits by 11\% and 23\%, independently of the mass, and the increase of the background by 7--8\%. Even in the most pessimistic case, the HL-LHC limits remain a factor of about 2 stronger than the Run~2 limits below 40~GeV.

\section{Comparison with other constraints}
\label{sec:constraints}

\begin{figure}[t]
\centering
\includegraphics[width=\columnwidth]{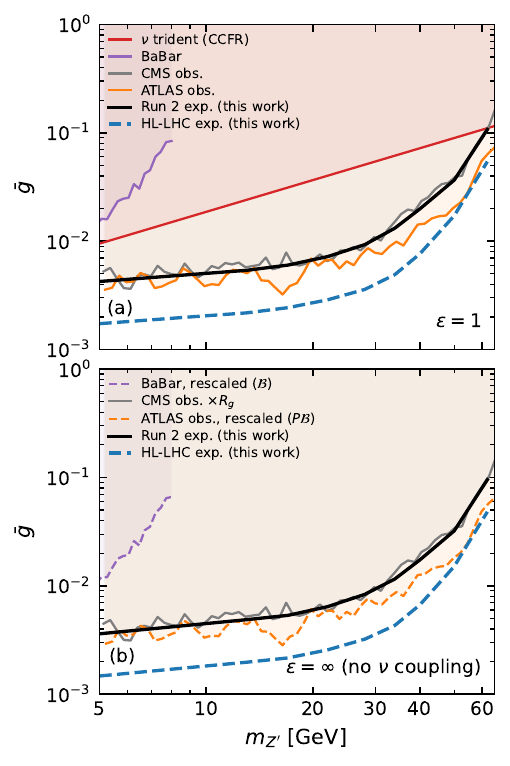}
\caption{Constraints on $\bar g$ as a function of $m_{Z^\prime}$ for (a) $\epsilon=1$ and (b) $\epsilon=\infty$. Shown are our expected limits for Run~2 (solid black) and for the HL-LHC (dashed blue), the observed limits of CMS~\cite{CMS:2018yxg}, ATLAS~\cite{ATLAS:2023fhg} and BaBar~\cite{BaBar:2016sci}, and the constraint from neutrino trident production~\cite{Mishra:1991bv, Altmannshofer:2016jzy}; shaded regions are excluded. In (b), the CMS limits are rescaled with the ratio $R_g$ of this work, and the ATLAS and BaBar limits (dashed) with the ratios of production cross sections and branching ratios; neutrino trident production does not constrain this case.}
\label{fig:money}
\end{figure}

Figure~\ref{fig:money}(a) compares our expected limits for $\epsilon=1$ with the observed limits of the CMS~\cite{CMS:2018yxg} and ATLAS~\cite{ATLAS:2023fhg} four-muon searches, of the BaBar search for $e^+e^-\to\mu^+\mu^-Z^\prime$ with $Z^\prime\to\mu^+\mu^-$~\cite{BaBar:2016sci}, and of neutrino trident production measured by CCFR~\cite{Mishra:1991bv}, as derived in Refs.~\cite{Altmannshofer:2014pba, Altmannshofer:2016jzy}. Our expected limits are consistent with the observed CMS limits within their statistical fluctuations. The ATLAS limits are stronger, by up to a factor of two above 30~GeV, owing to the larger integrated luminosity and the use of a neural network. Over the full mass range considered here, the four-muon searches are more sensitive than neutrino trident production, while the BaBar search constrains only masses below about 8~GeV, where it is weaker than the LHC searches. The HL-LHC projection of Sec.~\ref{sec:hllhc} extends the sensitivity by a factor of 2--2.7. We do not show constraints that require couplings of the $Z^\prime$ to quarks, such as $B_s$ mixing, since they depend on additional model assumptions.

The constraints depend on $\epsilon$ in different ways. Neutrino trident production requires a coupling of the $Z^\prime$ to muon neutrinos, which is proportional to $g_L$, and therefore vanishes for $\epsilon\to\infty$. In the BaBar search the muons are produced by a virtual photon, so that $P(\epsilon)=1$ and only the branching ratio changes. For the LHC searches, both the production cross section and the branching ratio change, as discussed in Sec.~\ref{sec:chiral}. Figure~\ref{fig:money}(b) shows the corresponding constraints for $\epsilon=\infty$. The CMS observed limits are rescaled with the ratio $R_g$ obtained in this work, which includes the change of the acceptance because our analysis reproduces the CMS selection. The ATLAS limits are rescaled with $[P(\epsilon)\,\mathcal B(\epsilon)/\mathcal B(1)]^{-1/2}$, and the BaBar limits with $[\mathcal B(\epsilon)/\mathcal B(1)]^{-1/2}$; these rescalings do not include possible changes of the acceptance, which reach 9\% at low masses for the CMS selection. For a $Z^\prime$ coupled only to right-handed leptons, the constraints from neutrino trident production no longer apply, while the searches with charged leptons remain sensitive, with limits on $\bar g$ that are 9--15\% stronger than for the vector model.

\section{Conclusions}
\label{sec:conclusions}

We have reinterpreted the CMS search for $Z\to\mu^+\mu^-Z^\prime$, $Z^\prime\to\mu^+\mu^-$ with $77.3~\mathrm{fb}^{-1}$ at $\sqrt s=13$~TeV for a $Z^\prime$ with the flavor structure of $L_\mu-L_\tau$ and independent couplings to left- and right-handed leptons, parametrized by $\epsilon=g_R/g_L$. For the vector model, our expected limits agree with those of CMS within 9\% for $m_{Z^\prime}=5$--62~GeV, and the fraction of signal events within the signal window agrees with the CMS values. The remaining difference at low masses is accounted for by the branching ratio and by a 10\% lower efficiency for muons with $p_T<10$~GeV, which reduces it to 3\% or less.

The dependence of the signal on $\epsilon$ separates into three factors. The dependence of the production cross section on $\epsilon$ arises only from the $Z$ couplings to left- and right-handed muons, and the cross section decreases by up to 15\% for right-handed-dominated couplings. The branching ratio to muons increases, at high masses from $1/3$ for the vector model to $1/2$ for $\epsilon\to\infty$, because the coupling to neutrinos decreases. The acceptance is the same as for the vector model for $\epsilon=-1$, which confirms the assumption used in Ref.~\cite{Fang:2026axial} to rescale the vector limits to the axial case, but decreases by up to 9\% at low masses for $|g_L|\neq|g_R|$, through the spin correlations between the emission and the decay of the $Z^\prime$. As a result, the expected limits on $\bar g$ are 7--15\% stronger for $\epsilon=3$, 7 and $\infty$ than for the vector model. For $\epsilon=-1$, the limit is 10\% stronger at 5~GeV, because of the branching ratio near the $\tau^+\tau^-$ threshold, and within 2\% of the vector limit above 13~GeV. Extending the rescaling with the ratio of branching ratios to $|g_L|\neq|g_R|$ gives limits that are too strong by up to 12\%; including the production cross section reduces this difference to about 2\% above 15~GeV and to 2--5\% between 5 and 13~GeV, where the acceptance changes.

For couplings dominated by right-handed leptons, the constraints from neutrino trident production are weakened, and vanish for $\epsilon=\infty$, while the searches with charged leptons remain sensitive. At the HL-LHC, with $3000~\mathrm{fb}^{-1}$ at $\sqrt s=14$~TeV and a background systematic uncertainty reduced by half, the sensitivity to $\bar g$ improves by a factor of 2.0--2.7, reaching $\bar g\simeq1.7\times10^{-3}$ at $m_{Z^\prime}=5$~GeV for $\epsilon=1$. The projection depends mainly on the background systematic uncertainty, and a 5--10\% lower muon efficiency due to pileup would weaken it by 11--23\%. 
The models with $\epsilon\neq\pm1$ require additional fermions to cancel the remaining gauge anomaly, and all models with $\epsilon\neq1$ require an extended sector to generate the muon and tau masses, as discussed in Appendix~\ref{app:eft}.

\section*{Data availability}
The expected limits for all benchmarks, the signal yields and the efficiencies are available from the author upon reasonable request.

\begin{acknowledgments}
I thank Teruki Kamon (Texas A\&M University) for his comments and suggestions throughout this work, and Eduardo Rojas (Universidad de Nari\~no) and Jos\'e David Alvarez-Ruiz (Universidad de Antioquia) for their guidance and many useful discussions. I thank the CMS group at Universidad de Antioquia for the training and methodology shared during my participation in the collaboration, to which I contributed Monte Carlo simulations. This experience was fundamental to independently design and develop the columnar analysis framework used in this work.

This work was partially supported by the Ministry of Science, Technology and Innovation of Colombia (Minciencias) under Call No.~860 de 2019. I acknowledge the support of the Faculty of Exact and Natural Sciences and the Institute of Physics of Universidad de Antioquia, and of Ghent University, where this work is part of my doctoral research.

The analysis code, simulations and results are the work of the author. Claude (Anthropic) was used as an assistant to edit and rewrite the English text. After using this tool, the author reviewed and edited the content as needed and takes full responsibility for the content of this paper.

\end{acknowledgments}

\appendix

\section{Theoretical consistency}
\label{app:eft}

With the charges of Eq.~\eqref{eq:lagrangian}, all gauge anomalies involving $U(1)^\prime$ cancel between the muon and tau families, except the mixed anomaly with two $U(1)^\prime$ bosons and one hypercharge boson. Summing over left-handed doublets and right-handed singlets of both families, its coefficient is
\begin{equation}
\mathcal A\big[U(1)_Y\,U(1)^{\prime\,2}\big]\propto 2\left(g_R^2-g_L^2\right)=4\bar g^2\,\frac{\epsilon^2-1}{\epsilon^2+1},
\label{eq:anomaly}
\end{equation}
which vanishes for $\epsilon=\pm1$. The vector ($\epsilon=1$) and axial ($\epsilon=-1$) models are therefore anomaly free with the SM fermions, while for $\epsilon=3$, 7 and $\infty$ additional fermions are required.

The SM Yukawa coupling $\bar L_\mu H\mu_R$ is invariant under $U(1)^\prime$ only for $g_L=g_R$. For $\epsilon\neq1$, the muon and tau masses must be generated in a different way, for example through the mixing with heavy vector-like leptons induced by a scalar that breaks $U(1)^\prime$. In such constructions the couplings of the light leptons to the $Z^\prime$ are chiral, and the heavy fermions can be chosen to cancel the anomaly of Eq.~\eqref{eq:anomaly}~\cite{Ellis:2017nrp, King:2017anf}. A renormalizable completion of the axial model is given in Ref.~\cite{Fang:2026axial}.

Below the masses of the heavy fermions, the benchmarks with $\epsilon\neq\pm1$ are described by an effective theory with an anomalous $U(1)^\prime$. Since the scalar that breaks $U(1)^\prime$ has a charge of order one and a vacuum expectation value of order $m_{Z^\prime}/\bar g$, the heavy fermions have masses $M\lesssim\sqrt{4\pi}\,m_{Z^\prime}/\bar g$, in the TeV range, which sets the cutoff of the effective theory. This is well below the maximal cutoff allowed by the mixed anomaly, which, following the arguments of Ref.~\cite[Sec.~4]{Preskill:1990fr}, is parametrically of order $64\pi^3 m_{Z^\prime}/(\bar g^2 g_Y|\mathcal A|)\sim10^{8}$~GeV for $m_{Z^\prime}=5$~GeV and $\bar g=10^{-2}$, where $\mathcal A$ is the coefficient of Eq.~\eqref{eq:anomaly} in units of $\bar g^2$. Integrating out the heavy fermions generates a Wess--Zumino term for the Goldstone boson of $U(1)^\prime$~\cite{Preskill:1990fr,DHoker:1984izu}, which induces effective interactions among the $Z^\prime$ and the hypercharge gauge boson.

These interactions~\cite{Anastasopoulos:2006cz} always involve two $Z^\prime$ bosons. Processes with a single $Z^\prime$ that are enhanced for anomalous currents, such as $Z\to Z^\prime\gamma$ or flavor-changing meson decays~\cite{Dror:2017ehi, Dror:2017nsg}, are therefore absent. The induced decay $Z\to Z^\prime Z^\prime$ is suppressed by the anomaly coefficient and by a loop factor, and its contribution to the four-muon final state is negligible compared with the signal for the couplings considered here.

Finally, we assume that there is no kinetic mixing between the $Z^\prime$ and the photon and no mass mixing between the $Z^\prime$ and the $Z$ at tree level. Kinetic mixing induced by muon and tau loops is proportional to the vector coupling $(g_L+g_R)/2$ and vanishes for $\epsilon=-1$. Loops also induce a $Z$--$Z^\prime$ mass mixing through the axial couplings, proportional to $m_\tau^2-m_\mu^2$, which does not vanish for $\epsilon=-1$ but is of order $10^{-8}$ for $\bar g=10^{-2}$. Both effects lead to couplings to electrons and quarks suppressed by a loop factor, which do not affect the four-muon signal.

\begin{figure}[t]
\centering
\includegraphics[width=\columnwidth]{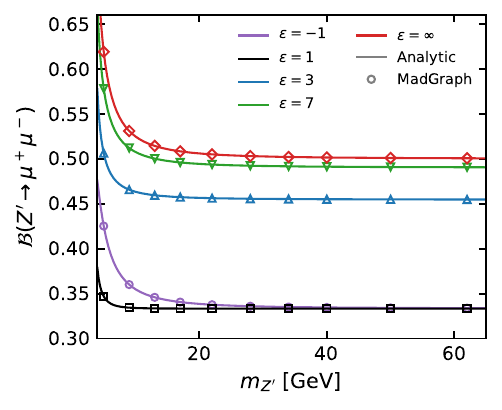}
\caption{$\mathcal B(Z^\prime\to\mu^+\mu^-)$ as a function of $m_{Z^\prime}$ for the benchmarks of Table~\ref{tab:benchmarks}, from Eqs.~\eqref{eq:gll}--\eqref{eq:gnn} (lines) and from \textsc{MadGraph5\_aMC@NLO} (markers). Above about 20~GeV the values for $\epsilon=-1$ and $\epsilon=1$ agree within 1.5\%.}
\label{fig:br_mass}
\end{figure}

Figure~\ref{fig:br_mass} compares the branching ratios computed from Eqs.~\eqref{eq:gll}--\eqref{eq:gnn} with those obtained with the model implementation of Sec.~\ref{sec:lagrangian}. The two agree within 0.03\% for all benchmarks and masses, which validates the implementation of the chiral couplings.

\section{Selection efficiencies}
\label{app:cutflow}

Table~\ref{tab:cutflow} gives the efficiency of each selection step for $m_{Z^\prime}=5$, 9, 13 and 40~GeV and the five benchmarks. The efficiencies are the same for all benchmarks within the statistical uncertainty, except for the veto on $m(\mu^+\mu^-)<4$~GeV and the signal window. For $\epsilon=3$, 7 and $\infty$ both efficiencies decrease at low masses, and the differences with respect to $\epsilon=1$ become smaller as the mass increases. At 40~GeV the efficiency of the veto still decreases by 1\%, while the total acceptance is the same for all benchmarks within the statistical uncertainty. For $\epsilon=-1$ all efficiencies agree with those of the vector model.

\begin{table}[t]
\caption{Efficiency of each selection step relative to the previous one for the five benchmarks. The first row includes the muon identification, kinematic and isolation requirements. The last row of each block gives the total acceptance $A(\epsilon)$, defined as the fraction of generated events within the signal window. The statistical uncertainty of each entry is below 0.005.}
\label{tab:cutflow}
\begin{ruledtabular}
\begin{tabular}{lccccc}
Selection step & \multicolumn{5}{c}{$\epsilon$} \\
 & $-1$ & $1$ & $3$ & $7$ & $\infty$ \\
\hline
\multicolumn{6}{c}{$m_{Z^\prime}=5$~GeV} \\
Muon selection and trigger & 0.902 & 0.903 & 0.904 & 0.904 & 0.902 \\
$\geq4\mu$, $p_T$ requirements & 0.452 & 0.452 & 0.455 & 0.456 & 0.455 \\
$Z_1$ and $Z_2$ candidates & 1.000 & 1.000 & 1.000 & 1.000 & 1.000 \\
$\Delta R>0.02$ & 0.999 & 0.999 & 0.999 & 0.999 & 0.999 \\
$m(Z_1)>12$~GeV & 1.000 & 1.000 & 1.000 & 1.000 & 1.000 \\
$80<m(4\mu)<100$~GeV & 0.919 & 0.916 & 0.919 & 0.918 & 0.919 \\
$m(\mu^+\mu^-)>4$~GeV & 0.886 & 0.890 & 0.868 & 0.859 & 0.857 \\
Signal window ($\pm2\%$) & 0.623 & 0.626 & 0.611 & 0.599 & 0.585 \\
Total, $A(\epsilon)$ & 0.207 & 0.208 & 0.200 & 0.195 & 0.189 \\
\hline
\multicolumn{6}{c}{$m_{Z^\prime}=9$~GeV} \\
Muon selection and trigger & 0.906 & 0.903 & 0.904 & 0.904 & 0.906 \\
$\geq4\mu$, $p_T$ requirements & 0.440 & 0.441 & 0.438 & 0.438 & 0.438 \\
$Z_1$ and $Z_2$ candidates & 0.999 & 1.000 & 1.000 & 1.000 & 0.999 \\
$\Delta R>0.02$ & 0.999 & 0.999 & 1.000 & 0.999 & 0.999 \\
$m(Z_1)>12$~GeV & 1.000 & 1.000 & 1.000 & 1.000 & 1.000 \\
$80<m(4\mu)<100$~GeV & 0.918 & 0.915 & 0.918 & 0.917 & 0.918 \\
$m(\mu^+\mu^-)>4$~GeV & 0.950 & 0.950 & 0.939 & 0.931 & 0.929 \\
Signal window ($\pm2\%$) & 0.601 & 0.608 & 0.595 & 0.590 & 0.590 \\
Total, $A(\epsilon)$ & 0.209 & 0.210 & 0.203 & 0.199 & 0.200 \\
\hline
\multicolumn{6}{c}{$m_{Z^\prime}=13$~GeV} \\
Muon selection and trigger & 0.913 & 0.916 & 0.915 & 0.917 & 0.916 \\
$\geq4\mu$, $p_T$ requirements & 0.440 & 0.442 & 0.441 & 0.443 & 0.439 \\
$Z_1$ and $Z_2$ candidates & 1.000 & 1.000 & 1.000 & 1.000 & 0.999 \\
$\Delta R>0.02$ & 1.000 & 1.000 & 1.000 & 1.000 & 1.000 \\
$m(Z_1)>12$~GeV & 1.000 & 1.000 & 1.000 & 1.000 & 1.000 \\
$80<m(4\mu)<100$~GeV & 0.917 & 0.923 & 0.915 & 0.917 & 0.916 \\
$m(\mu^+\mu^-)>4$~GeV & 0.968 & 0.970 & 0.961 & 0.957 & 0.955 \\
Signal window ($\pm2\%$) & 0.610 & 0.616 & 0.604 & 0.589 & 0.589 \\
Total, $A(\epsilon)$ & 0.217 & 0.223 & 0.214 & 0.210 & 0.207 \\
\hline
\multicolumn{6}{c}{$m_{Z^\prime}=40$~GeV} \\
Muon selection and trigger & 0.966 & 0.968 & 0.966 & 0.967 & 0.967 \\
$\geq4\mu$, $p_T$ requirements & 0.538 & 0.538 & 0.539 & 0.540 & 0.534 \\
$Z_1$ and $Z_2$ candidates & 1.000 & 1.000 & 1.000 & 1.000 & 1.000 \\
$\Delta R>0.02$ & 1.000 & 1.000 & 1.000 & 1.000 & 1.000 \\
$m(Z_1)>12$~GeV & 1.000 & 1.000 & 1.000 & 1.000 & 1.000 \\
$80<m(4\mu)<100$~GeV & 0.887 & 0.889 & 0.886 & 0.888 & 0.884 \\
$m(\mu^+\mu^-)>4$~GeV & 0.982 & 0.981 & 0.977 & 0.972 & 0.970 \\
Signal window ($\pm2\%$) & 0.253 & 0.253 & 0.257 & 0.258 & 0.264 \\
Total, $A(\epsilon)$ & 0.114 & 0.115 & 0.116 & 0.116 & 0.117 \\
\end{tabular}
\end{ruledtabular}
\end{table}

\bibliography{references}

\end{document}